\documentclass[pra,preprint,superscriptaddress]{revtex4}
\usepackage{amsmath}
\usepackage{graphicx}
\usepackage{hyperref}
\usepackage{theorem}

\theorembodyfont{\normalfont}
\newtheorem{theorem}{Theorem}
\newtheorem{definition}[theorem]{Definition}
\newtheorem{corollary}[theorem]{Corollary}

\begin{document}

YITP-18-05
\title{Merlin-Arthur with efficient quantum Merlin
and quantum supremacy for
the second level of the Fourier hierarchy
}
\author{Tomoyuki Morimae}
\email{tomoyuki.morimae@yukawa.kyoto-u.ac.jp}
\affiliation{Yukawa Institute for Theoretical Physics,
Kyoto University, Kitashirakawa Oiwakecho, Sakyo-ku, Kyoto
606-8502, Japan}
\affiliation{Department of Computer Science, 
Gunma University, 1-5-1 Tenjincho, Kiryu,
Gunma, 376-0052, Japan}
\affiliation{JST, PRESTO, 4-1-8 Honcho, Kawaguchi, Saitama,
332-0012, Japan}
\author{Yuki Takeuchi}
\email{takeuchi.yuki@lab.ntt.co.jp}
\affiliation{Graduate School of Engineering Science, 
Osaka University, Toyonaka, Osaka 560-8531, Japan}
\affiliation{NTT Communication Science Laboratories, NTT
Corporation, 3-1 Morinosato Wakamiya, Atsugi, Kanagawa
243-0198, Japan}
\author{Harumichi Nishimura}
\email{hnishimura@is.nagoya-u.ac.jp}
\affiliation{Graduate School of Informatics, Nagoya University,
Furocho, Chikusaku, Nagoya, Aichi, 464-8601, Japan}

\date{\today}
\begin{abstract}
We introduce a simple sub-universal quantum computing model,
which we call 
the Hadamard-classical circuit with one-qubit (HC1Q) model.
It consists of a classical reversible circuit
sandwiched by two layers of Hadamard gates,
and therefore it is in the second level of the Fourier hierarchy~\cite{FH}.
We show that 
output probability distributions of the HC1Q model
cannot be classically efficiently sampled
within a multiplicative error 
unless the polynomial-time hierarchy
collapses to the second level.
The proof technique is different from
those used for previous sub-universal models, such as IQP, Boson Sampling,
and DQC1, and therefore the technique itself
might be useful for finding other sub-universal models that
are hard to classically simulate.
We also study the classical verification of quantum computing
in the second level of the Fourier hierarchy.
To this end,
we define a promise problem, which we call the probability distribution 
distinguishability with maximum norm (PDD-Max). It is a promise
problem to decide whether output probability distributions of two
quantum circuits are far apart or close.
We show that PDD-Max
is BQP-complete, but if the two circuits
are restricted to some types in the second level of the Fourier hierarchy,
such as the HC1Q model or the IQP model,
PDD-Max has a Merlin-Arthur system
with quantum polynomial-time Merlin and classical probabilistic
polynomial-time Arthur.
\end{abstract}

%\pacs{}
\maketitle

\section{Introduction}
\subsection{Quantum supremacy of the HC1Q model}
The Fourier hierarchy~\cite{FH} is a hierarchy of
restricted quantum circuits.
The $k$th level of the Fourier hierarchy, 
FH$_k$, 
is the class of 
quantum circuits 
with $k$ layers of Hadamard gates and all other gates
preserving the computational basis.
The second level, FH$_2$, is the most important,
because circuits in FH$_2$ are used in many algorithms,
such as
Simon's algorithm~\cite{Simon} and 
Shor's factoring algorithm~\cite{Shor}.
The instantaneous quantum polynomial-time (IQP) 
model~\cite{BJS,BMS} is also in FH$_2$:
\begin{definition}[IQP]
The IQP model on $N$ qubits is the following quantum computing model:
\begin{itemize}
\item[1.]
The initial state is $|0^N\rangle$.
\item[2.]
The unitary $H^{\otimes N}DH^{\otimes N}$ is applied.
Here $H$ is the Hadamard gate,
and $D$ is a quantum circuit
consisting of only $Z$-diagonal gates, 
such as $Z$, $CZ$, $CCZ$, and $e^{i\theta Z}$, etc.
\item[3.]
All qubits are measured in the computational basis.
\end{itemize}
\end{definition}
The IQP model is a well-known example of 
sub-universal quantum computing models
whose output probability distributions cannot be classically efficiently
sampled unless the polynomial-time hierarchy 
collapses.
Since a collapse of the polynomial-time hierarchy is considered
as highly unlikely 
in computer science, it shows the hardness of classically
simulating 
the IQP model.
Other sub-universal models that exhibit similar quantum
supremacy are also known, such as
the depth four model~\cite{TD},
the Boson sampling model~\cite{AA},
the DQC1 model~\cite{KL,MFF,M,Kobayashi,KobayashiICALP},
the Fourier sampling model~\cite{BU},
the conjugated Clifford model~\cite{AFK},
and the random circuit model~\cite{random}.

\begin{figure}[htbp]
\begin{center}
\includegraphics[width=0.6\textwidth]{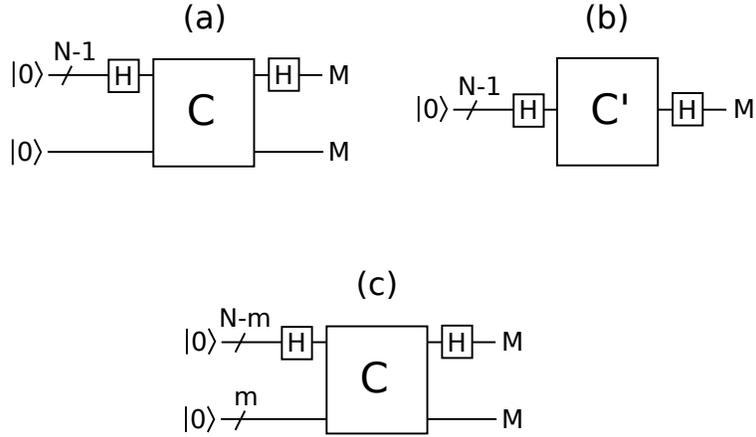}
\end{center}
\caption{
(a) The HC1Q model.
$M$ is the computational-basis measurement.
A single line with the slash represents a set of $N-1$ qubits.
The Hadamard gate $H$ is applied on each qubit.
The measurement $M$ is done on each qubit.
(b) The circuit similar to HC1Q model, but the last $|0\rangle$
qubit is removed.
(c) The generalization of the HC1Q model, which we call the HC$m$Q model.
} 
\label{simple2}
\end{figure}

The first main result of the present paper is
to add another simple model in FH$_2$
to the above list of sub-universal models 
that exhibit quantum supremacy.
We define the Hadamard-classical circuit with one-qubit (HC1Q) model
as follows:
\begin{definition}[HC1Q]
The Hadamard-classical circuit with one-qubit (HC1Q) model
on $N$ qubits
is the following quantum computing model (see Fig.~\ref{simple2}(a)):  
\begin{itemize}
\item[1.]
The initial state is 
$|0^N\rangle$.
\item[2.]
$H^{\otimes (N-1)}\otimes I$ is applied.
\item[3.]
A polynomial-time uniformly-generated classical
reversible circuit 
\begin{eqnarray*}
C:\{0,1\}^N\ni w
\mapsto C(w)\in\{0,1\}^N
\end{eqnarray*}
is applied ``coherently".
\item[4.]
$H^{\otimes (N-1)}\otimes I$ is applied.
\item[5.]
All qubits are measured in the computational basis.
\end{itemize}
\end{definition}

We show that
the HC1Q model is universal with a postselection.
More precisely, we show the following:

\begin{theorem}
\label{supremacy}
Let $U$ be a polynomial-time uniformly-generated
quantum circuit acting on $n$ qubits
that consists of $poly(n)$ number of
Hadamard gates and classical reversible gates, such as 
$X$, CNOT, and Toffoli.
Then from $U$ we can efficiently construct
an HC1Q circuit on $N=poly(n)$ qubits
such that
a postselection on some output qubits 
generates the state $U|0^n\rangle$.
\end{theorem}
Since Hadamard plus classical gates are universal,
Theorem~\ref{supremacy} 
shows that HC1Q model is universal with a postselection.
A proof of the theorem is given in Sec.~\ref{Sec:supremacy}.
The proof is based on
a new idea that is different from those used in the previous
results~\cite{TD,BJS,AA,MFF}.
Proofs for the depth-four model~\cite{TD},
the Boson sampling model~\cite{AA}, and the IQP model~\cite{BJS}
use gadgets that insert gates in a post hoc way by using 
postselections. 
(For example, the IQP model is not universal because Hadamard
gates can be applied only in the first layer and the last layer,
but the so-called Hadamard gadget~\cite{BJS} 
enables realizations of Hadamard gates
in any place of the IQP model by using postselections.)
The proof for the DQC1 model~\cite{MFF} uses a postselection
to initialize the maximally-mixed state $I^{\otimes n}/2^n$
to the pure state $|0^n\rangle$.
As is explained in Sec.~\ref{Sec:supremacy},
our proof is different from those:
from $U$, we first construct a polynomial-time non-deterministic algorithm.
From the polynomial-time non-deterministic algorithm,
we next construct a classical deterministic polynomial-time circuit $C$.
We finally show that the HC1Q circuit 
$(H^{\otimes N-1}\otimes I)C(H^{\otimes N-1}\otimes I)$ 
can generate
$U|0^n\rangle$ with a postselection.
This idea itself seems to be useful for finding
other new sub-universal models that exhibit
quantum supremacy.
We do not know relations between our technique and previous ones.

By using the arguments of Ref.~\cite{BJS,MFF,AA},
we obtain the following corollary of Theorem~\ref{supremacy},
which is quantum supremacy of the HC1Q model:

\begin{corollary}
Output probability
distributions of the HC1Q model 
cannot be classically efficiently sampled within a multiplicative
error $\epsilon<1$ unless the polynomial-time hierarchy
collapses to the third level.
\end{corollary}
Here, we say that a probability distribution $\{p_z\}_z$ is classically
efficiently sampled within a multiplicative error $\epsilon$ if
there exists a classical probabilistic polynomial-time 
algorithm such that
\begin{eqnarray*}
|p_z-q_z|\le\epsilon p_z
\end{eqnarray*}
for all $z$, where $q_z$ is the probability that the algorithm
outputs $z$.

The corollary demonstrates
an interesting ``phase transition"
between classical and quantum.
To see it, let us consider the circuit of Fig.~\ref{simple2}(b),
which is obtained by removing the last $|0\rangle$ qubit of
the HC1Q model (Fig.~\ref{simple2}(a)).
Here,
\begin{eqnarray*}
C':\{0,1\}^{N-1}\ni w\mapsto C'(w)\in\{0,1\}^{N-1}
\end{eqnarray*}
is a polynomial-time uniformly-generated classical reversible circuit. 
The circuit of Fig.~\ref{simple2}(b) 
is trivially classically simulatable, since
$C'$ is a permutation on $\{0,1\}^{N-1}$ and therefore
\begin{eqnarray*}
C'|+\rangle^{\otimes (N-1)}=\frac{1}{\sqrt{2^{N-1}}}
\sum_{x\in\{0,1\}^{N-1}}|C'(x)\rangle
=|+\rangle^{\otimes (N-1)},
\end{eqnarray*}
where $|+\rangle\equiv(|0\rangle+|1\rangle)/\sqrt{2}$.
Our result therefore suggests that
the addition of a single $|0\rangle$ qubit to the trivial circuit
of Fig.~\ref{simple2}(b) changes
its complexity dramatically.

The third level collapse of the polynomial-time hierarchy
for the depth-four model,
the Boson sampling model, the IQP model, and the DQC1 model 
can be improved
to the second level collapse~\cite{Kobayashi,KobayashiICALP}.
In Sec.~\ref{Sec:supremacy2},
we show that the same improvement is possible
for the HC1Q model:

\begin{theorem}
\label{supremacy2}
Output probability
distributions of the HC1Q model
cannot be classically efficiently sampled within a multiplicative
error $\epsilon<1$ unless the polynomial-time hierarchy
collapses to the second level.
\end{theorem}

A natural question is whether the above supremacy results still
hold when the number of qubits measured is much reduced.
It is known that for the $N$-qubit IQP model
probability distributions of measurement results on
$O(\log(N))$ number of qubits can be classically exactly sampled
in polynomial time~\cite{BJS}.
In Sec.~\ref{Sec:log_classical} we show an analogous result for 
the HC1Q model:
\begin{theorem}
\label{simulatability}
Probability distributions of
measurement results on
$O(\log(N))$ number of qubits in  
the HC1Q model 
can be classically
sampled in polynomial time within a $1/poly(N)$ L1-norm error.
\end{theorem}
It is possible to extend this result
to a generalized version of the HC1Q model, which
we call the HC$m$Q model:
\begin{definition}[HC$m$Q]
The Hadamard-classical circuit with $m$-qubit (HC$m$Q) model
on $N$ qubits
is the following quantum computing model (see Fig.~\ref{simple2}(c)):  
\begin{itemize}
\item[1.]
The initial state is 
$|0^N\rangle$.
\item[2.]
$H^{\otimes (N-m)}\otimes I^{\otimes m}$ is applied.
\item[3.]
A polynomial-time uniformly-generated classical
reversible circuit 
\begin{eqnarray*}
C:\{0,1\}^N\ni w
\mapsto C(w)\in\{0,1\}^N
\end{eqnarray*}
is applied ``coherently".
\item[4.]
$H^{\otimes (N-m)}\otimes I^{\otimes m}$ is applied.
\item[5.]
All qubits are measured in the computational basis.
\end{itemize}
\end{definition}

\begin{theorem}
\label{simulatability2}
Probability distributions of
measurement results on
$O(\log(N))$ number of qubits in  
the HC$m$Q model 
can be classically
sampled in polynomial time within a $1/poly(N)$ L1-norm error.
\end{theorem}
Its proof is omitted since it is similar to that of 
Theorem~\ref{simulatability}.

At this moment, we do not know which sub-universal model is the most promising
for experimental realizations, but the HC1Q model should be useful
for certain experimental setups due to its simple structure.

\subsection{The verification of quantum computing in the
second level of the Fourier hierarchy}
In this paper, we also study the classical verification of quantum
computing.
It is a long-standing open problem
whether quantum computing is classically verifiable.
More precisely, it is open whether
any problem $L$ in BQP has an
interactive proof system with a quantum polynomial-time
prover and a classical probabilistic polynomial-time verifier. 
\begin{definition}
We say that a problem $L$ has
an interactive proof system with a quantum
polynomial-time prover if there exists a classical
probabilistic polynomial-time verifier such that
\begin{itemize}
\item
If $x\in L$ then there exists a quantum polynomial-time prover
such that the verifier accepts
with probability at least $2/3$.
\item
If $x\notin L$ then for any prover
the verifier accepts
with probability at most $1/3$.
\end{itemize}
\end{definition}
(Note that in this definition, 
the prover is in quantum polynomial-time 
only for yes instances, i.e., when the prover is honest.
For no instances,
the computational power of the malicious prover is unbounded.)

Answering the open question is 
important not only for practical applications of cloud quantum computing
but also for foundations of computer science
and quantum physics~\cite{AharonovVazirani}.
In fact, several partial solutions to the open problem
have been obtained. They are categorized into the following four types: 
\begin{itemize}
\item[1.]
Several verification protocols~\cite{MNS,posthoc} 
and verifiable blind quantum computing 
protocols~\cite{FK,Aharonov,HM,Broadbent,Andru_review}
demonstrate that if the verifier has a weak quantum ability,
such as preparations or measurements of single-qubit quantum states,
any BQP problem can be verified with a 
quantum polynomial-time prover.
\item[2.]
If multiple entangling 
quantum polynomial-time provers
who are not communicating with each other are 
allowed, any BQP problem is verified with
a classical polynomial-time verifier~\cite{MattMBQC,RUV,Ji}.
\item[3.]
Since BQP is contained in IP~\cite{defIP},
a natural approach to the open problem is to restrict the prover
of IP to quantum polynomial-time when the problem is in BQP.
In fact, recently, a step in this line
has been obtained in Ref.~\cite{AharonovGreen}.
The authors of Ref.~\cite{AharonovGreen} 
have constructed a new interactive proof system
that verifies the value of the trace of operators
with a postBQP prover 
and a classical polynomial-time verifier.
\item[4.]
It has been shown recently that the classical verification
of quantum computing is possible 
if a certain problem is assumed to be hard for quantum 
computing~\cite{Mahadev}.
\end{itemize}

Actually, the answer to the open problem is unconditionally yes
if we consider specific BQP problems.
For example, it is known that
the recursive Fourier sampling~\cite{BV}
has an interactive proof system with a 
quantum polynomial-time prover and
a classical polynomial-time verifier
who communicate polynomial number of 
messages~\cite{MattFourier}.
Furthermore,
it has been shown recently that calculating orders of solvable
groups has an interactive proof system with a 
quantum polynomial-time prover
and a classical polynomial-time 
verifier who exchange two or three 
messages~\cite{LeGall}.
Finally,
it was suggested in Ref.~\cite{Tommaso}
that a problem of deciding whether there exist some results
that occur with high probability or not for circuits 
in FH$_2$ has
a Merlin-Arthur system with quantum polynomial-time
Merlin.
\begin{definition}
We say that a problem $L$ has
a Merlin-Arthur system with a quantum
polynomial-time Merlin if $L$ has an interactive proof system
with a quantum polynomial-time prover (Merlin) and
a classical probabilistic polynomial-time verifier (Arthur)
where only a single 
message transmission is done from the prover to the verifier.
\end{definition}

The second main result of the present paper is to introduce
another problem in BQP that is classically verifiable.
More precisely, we define the following promise problem that we call
Probability Distribution Distinguishability
with Maximum Norm (PDD-Max):

\begin{definition}[PDD-Max]
Given a classical description of
two quantum circuits $U_1$ and $U_2$ acting on $N$ qubits
that consist of $poly(N)$ number of elementary gates,
and parameters $a$ and $b$ such that $0\le b<a\le1$ and
$a-b\ge1/poly(N)$,
decide
\begin{itemize}
\item
YES:
there exists $z\in\{0,1\}^N$ such that $|p_z-q_z|\ge a$.
\item
NO: for any $z\in\{0,1\}^N$, $|p_z-q_z|\le b$.
\end{itemize}
Here, 
$p_z\equiv |\langle z|U_1|0^N\rangle|^2$
and
$q_z\equiv |\langle z|U_2|0^N\rangle|^2$.
\end{definition}

We first show that PDD-Max exactly characterizes the power of
BQP: 
\begin{theorem}
\label{BQPcomplete}
PDD-Max is BQP-complete (under polynomial-time many-one reduction).
\end{theorem}
Its proof is given in Sec.~\ref{Sec:BQPcomp}.
(Note that if two circuits are classical, PDD-Max is BPP-complete;
we have only to replace the controlled-$H$ gate in Fig.~\ref{complete}
with a controlled-random-bit-flip gate.)

We next show that if $U_1$ and $U_2$ of
PDD-Max are restricted to the HC1Q model,
PDD-Max is classically verifiable:

\begin{theorem}
\label{restriction_HC1Q}
If two circuits
$U_1$ and $U_2$ are in the form of the HC1Q model,
PDD-Max has a Merlin-Arthur system with quantum polynomial-time Merlin.
\end{theorem}
The proof of Theorem~\ref{restriction_HC1Q}
is given in Sec.~\ref{Sec:FH_2}.
A similar proof holds for the IQP model and the 
HC$m$Q model (including the Simon-type circuits)
(see Sec.~\ref{Sec:discussion1}).

These results demonstrate that if we restrict the circuits of
PDD-Max to the form of the HC1Q model,
it is another example of problems
that is classically verifiable.
These results also suggest that such a restriction of 
PDD-Max 
is not BQP-hard, since BQP is not believed to be 
in MA~\cite{defMA}.

The classical verifiability of PDD-Max for restricted circuits
does not seem to be directly related to the classical
verification of quantum supremacy, such as the verification of the IQP model,
but it is an important future research subject to explore
any relation between them.

\subsection{Preliminary}
In this paper, we often use the following well known inequality:

\begin{theorem}[Chernoff-Hoeffding bound]
Let $X_1,...,X_T$ be identically and independently distributed
real random variables with
$|X_i|\le1$ for every $i=1,...,T$. Then 
\begin{eqnarray*}
{\rm Pr}\Big[
\Big|\frac{1}{T}\sum_{i=1}^TX_i-E(X_i)\Big|\le \epsilon\Big]
\ge1-2e^{-\frac{T\epsilon^2}{2}}.
\end{eqnarray*}
\end{theorem}

In the following sections, we give proofs of theorems.
In the last section,
Sec.~\ref{Sec:discussion}, we provide some discussions.

\section{Proof of Theorem~\ref{supremacy}}
\label{Sec:supremacy}
In this section, we show Theorem~\ref{supremacy}.
We are given a unitary operator 
\begin{eqnarray*}
U=u_t...u_2u_1 
\end{eqnarray*}
acting on $n$ qubits, 
where $u_i$ is the Hadamard gate $H$ or a classical gate for
all $i=1,2,...,t$.
From $U$, we define
\begin{eqnarray*}
U'\equiv u_{t+n}...u_{t+1}U,
\end{eqnarray*}
where $u_{t+i}$ is $H$ acting on $i$th qubit ($i=1,2,...,n$).

\begin{figure}[htbp]
\begin{center}
\includegraphics[width=0.5\textwidth]{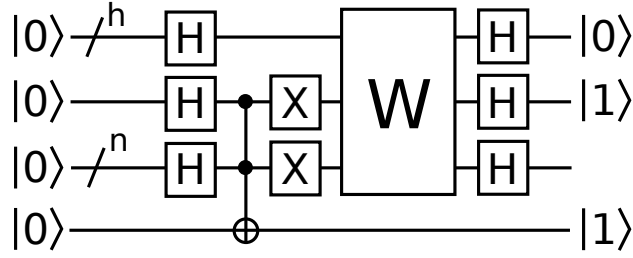}
\end{center}
\caption{
The quantum circuit with a postselection.
The first $h$ qubits, the $(h+1)$th qubit, and the last qubit
are postselected to $|0^h\rangle$, $|1\rangle$, and
$|1\rangle$, respectively.
} 
\label{proof2}
\end{figure}

The outline of our proof is as follows.
We first construct a 
polynomial-time non-deterministic 
algorithm from $U'$. We next define 
a polynomial-time quantum unitary operator 
$W$, which consists of only classical gates, 
from the polynomial-time non-deterministic algorithm.
We finally show that the HC1Q model of Fig.~\ref{proof2}
that uses $W$ is universal with
a postselection.

Now let us consider the following polynomial-time
non-deterministic algorithm.
\begin{itemize}
\item[0.]
The state of the register 
is represented by
$(s,z)$, where 
$s\in\{0,1\}$ and $z\in\{0,1\}^n$.
\item[1.]
The initial state of the register is 
$(s=0,z=0^n)$.
\item[2.]
Repeat the following for $i=1,...,t+n$.
\begin{itemize}
\item[2-a.]
If $u_i$ is a classical gate (such as $X$, CNOT, or Toffoli)
whose corresponding action on $\{0,1\}^n$ is
\begin{eqnarray*}
g:\{0,1\}^n\ni z\mapsto g(z)\in\{0,1\}^n,
\end{eqnarray*}
update the state of the register as
\begin{eqnarray*}
(s,z)
\to 
(s,g(z)).
\end{eqnarray*}
%For example, if $G$ is $CCX$ acts on $j,k,l$th bits with the target $l$th,
%$CCX(z)=z_1,...,z_j,...,z_k,...,z_l\oplus z_jz_k,...,z_n$.
\item[2-b.]
If $u_i$ is $H$ acting on $j$th qubit,
update the state of the register in the following
non-deterministic way: 
\begin{eqnarray*}
(s,z)
\to
\left\{
\begin{array}{ll}
(s,z_1,...,z_{j-1},0,z_{j+1},...,z_n)\\
(s\oplus z_j,z_1,...,z_{j-1},1,z_{j+1},...,z_n).
\end{array}
\right.
\end{eqnarray*}
\end{itemize}
\end{itemize}

Let us define $h$ by
\begin{eqnarray*}
h\equiv\big|\{i\in\{1,2,...,t+n\}~|~u_i=H\}\big|,
\end{eqnarray*}
i.e., $h$ is the number of $H$ appearing in $U'$.
The above polynomial-time non-deterministic algorithm
does the non-deterministic transition $h$ times, and therefore 
the algorithm has $2^h$ computational paths.
We label each path by an $h$-bit string $y\in\{0,1\}^h$.
We write the final state of the register for the path $y\in\{0,1\}^h$
by
$(s(y),z(y))$,
where 
$s(y)\in\{0,1\}$
and 
$z(y)\in\{0,1\}^n$.
Then, we obtain 
\begin{eqnarray}
U'|0^n\rangle=\frac{1}{\sqrt{2^h}}
\sum_{y\in\{0,1\}^h}(-1)^{s(y)}|z(y)\rangle.
\label{ahoaho}
\end{eqnarray}
(To help the readers understand the derivation of
Eq.~(\ref{ahoaho}), we provide a simple example in 
Appendix~\ref{app:example}.)

Note that, for each $y$, we can calculate $s(y)$ and $z(y)$ in classical
polynomial-time. 
Let us define the polynomial-time unitary operator $W$ by
\begin{eqnarray*}
W(|y\rangle\otimes|0\rangle\otimes|0^n\rangle)
=|y\rangle\otimes|s(y)\rangle\otimes|z(y)\rangle.
\end{eqnarray*}
It is easy to see that $W$ can be constructed with only classical
gates.
Now we show that the state 
$H^{\otimes n}U'|0^n\rangle=U|0^n\rangle$ 
can be generated
by the HC1Q model of Fig.~\ref{proof2} that uses $W$ with a postselection.
In Fig.~\ref{proof2}, the state immediately before the postselection is
\begin{widetext}
\begin{eqnarray}
\frac{1}{\sqrt{2^{n+h+1}}}
\sum_{y\in\{0,1\}^h}(H^{\otimes h}|y\rangle)\otimes
(H|s(y)\rangle)\otimes(H^{\otimes n}|z(y)\rangle)
\otimes|1\rangle
+|\psi\rangle\otimes|0\rangle,
\label{finalstate}
\end{eqnarray}
\end{widetext}
where $|\psi\rangle$ is a certain $(h+n+1)$-qubit state
whose detail is irrelevant~\cite{keep}.
After the postselection, the state becomes
\begin{eqnarray*}
\frac{1}{\sqrt{2^h}}
\sum_{y\in\{0,1\}^h}
(-1)^{s(y)}H^{\otimes n}|z(y)\rangle
=H^{\otimes n}U'|0^n\rangle
=U|0^n\rangle.
\end{eqnarray*}
Hence, we have shown that the HC1Q model of Fig.~\ref{proof2}
with a postselection
can generate $U|0^n\rangle$
for any unitary $U$ that consists of Hadamard and classical gates, 
which means that
the HC1Q model is universal
with a postselection.

\section{Proof of Theorem~\ref{supremacy2}}
\label{Sec:supremacy2}
%It was shown in Refs.~\cite{BJS,AA,MFF}
%that, by using the relation
%${\rm postBQP}={\rm PP}$~\cite{AaronsonpostBQP}, 
%if a sub-universal model becomes universal
%with a postselection,
%its output probability distribution cannot be classically
%efficiently sampled within a multiplicative error $\epsilon<1$
%unless the polynomial-time hierarchy collapses to the third 
%level.
%The result of the previous subsection therefore shows that
%the output probability distribution of the HC1Q model
%cannot be classically efficiently sampled within a multiplicative error
%$\epsilon<1$ unless the polynomial-time hierarchy collapses to the
%third level.

It was shown in Refs.~\cite{Kobayashi,KobayashiICALP} 
that the third level collapse
of the polynomial-time hierarchy for most of
the sub-universal models (including the depth-four model~\cite{TD},
the Boson sampling model~\cite{AA},
the IQP model~\cite{BJS}, and the DQC1 model~\cite{MFF})
is improved to the second level collapse.
The idea is to use the class NQP~\cite{NQP}
in stead of postBQP. 
NQP is a quantum version of NP, and
defined as follows:

\begin{definition}[NQP] A promise problem
$A=(A_{yes},A_{no})$ is in NQP if and only if there exists
a polynomial-time uniformly-generated family $\{V_x\}_x$ of
quantum circuits such that
if $x\in A_{yes}$ then $p_{acc}>0$,
and
if $x\in A_{no}$ then $p_{acc}=0$.
Here $p_{acc}$ is the acceptance probability of $V_x$.
\end{definition}

If we remember the following definition of NP, it is clear that
NQP is a quantum version of NP.

\begin{definition}[NP]
\label{defNP}
A promise problem
$A=(A_{yes},A_{no})$ is in NP if and only if there exists
a classical polynomial-time probabilistic
algorithm 
such that
if $x\in A_{yes}$ then $p_{acc}>0$,
and
if $x\in A_{no}$ then $p_{acc}=0$.
Here $p_{acc}$ is the acceptance probability of the classical algorithm.
\end{definition}

By using a similar argument of Refs.~\cite{Kobayashi,KobayashiICALP},
we now show that the collapse of the polynomial-time
hierarchy to the third level for the HC1Q model can be
improved to the second level.
Let $A=(A_{yes},A_{no})$ be a promise problem in NQP. 
Then there exists a polynomial-time uniformly-generated 
family $\{V_x\}_x$ of
quantum circuits that satisfies the above definition of NQP.
The outline of our proof is as follows:
First, from $V_x$, we construct an HC1Q circuit. Second, we assume that
its output probability distribution is classically efficiently sampled
within a multiplicative error $\epsilon<1$. Finally,
we show that then $A$ is in NP, which means 
${\rm NQP}\subseteq{\rm NP}$.
It collapses the polynomial-time hierarchy to the second level, since
\begin{eqnarray*}
{\rm PH}\subseteq{\rm BP}\cdot{\rm coC}_={\rm P}
={\rm BP}\cdot{\rm NQP}={\rm BP}\cdot {\rm NP}={\rm AM}.
\end{eqnarray*}

Now let us give a more precise proof.
In Sec.~\ref{Sec:supremacy},
we have constructed the
HC1Q circuit of Fig.~\ref{proof2} from a given unitary $U$.
Let us repeat the same argument by replacing
$U$ with $V_x$. 
In stead of doing the postselection on the state of 
Eq.~(\ref{finalstate}),
let us do the projective measurement
$\{\Lambda,I^{\otimes h+n+2}-\Lambda\}$
on it,
where
\begin{eqnarray*}
\Lambda\equiv
|0\rangle\langle 0|^{\otimes h}\otimes
|1\rangle\langle1|\otimes |0\rangle\langle0|\otimes I^{\otimes n-1}\otimes
|1\rangle\langle1|.
\end{eqnarray*}
The probability $p_\Lambda$ of obtaining $\Lambda$ is
\begin{eqnarray*}
p_\Lambda=\frac{\langle0^n|V_x^\dagger
(|0\rangle\langle0|\otimes I^{\otimes n-1})V_x|0^n\rangle}{2^{h+n+2}}.
\end{eqnarray*}
By the definition of NQP,
\begin{eqnarray*}
\langle0^n|V_x^\dagger
(|0\rangle\langle0|\otimes I^{\otimes n-1})V_x|0^n\rangle>0
\end{eqnarray*}
when $x\in A_{yes}$ and
\begin{eqnarray*}
\langle0^n|V_x^\dagger
(|0\rangle\langle0|\otimes I^{\otimes n-1})V_x|0^n\rangle=0
\end{eqnarray*}
when $x\in A_{no}$.
It means that $p_\Lambda>0$ when $x\in A_{yes}$ and
$p_\Lambda=0$ when $x\in A_{no}$.

Assume that $p_\Lambda$ is classically efficiently sampled
within a multiplicative error $\epsilon<1$. 
It means that
there exists a classical polynomial-time probabilistic algorithm
that outputs $0$ or $1$ such that
\begin{eqnarray*}
|p_\Lambda-q_0|\le \epsilon p_\Lambda,
\end{eqnarray*}
where $q_0$ is the probability that
the classical algorithm outputs 0 (accepts).
Then, we can show that $A$ is in NP. In fact,
if $x\in A_{yes}$ then
\begin{eqnarray*}
q_0\ge (1-\epsilon)p_\Lambda>0.
\end{eqnarray*}
If $x\in A_{no}$ then
\begin{eqnarray*}
q_0\le (1+\epsilon)p_\Lambda=0.
\end{eqnarray*}
According to Definition~\ref{defNP}, $A$ is therefore in NP.

\section{Proof of Theorem~\ref{simulatability}}
\label{Sec:log_classical}
In this section, we show Theorem~\ref{simulatability}.
For simplicity, we assume that the first
$k$ qubits are measured, where
$k=O(\log N)$.
Generalizations to other
$k$ qubits are the same.

The outline of our proof is as follows.
We first construct a probability distribution $\{q_z\}_z$
that can be calculated in classical polynomial time, and
is close to $\{p_z\}_z$ within a $1/poly$ L1-norm.
We then show that we can sample $\{q_z\}_z$ in classical polynomial time.

Let us consider the circuit of Fig.~\ref{simple2} (a).
As is shown in Appendix~\ref{app2},
the probability of obtaining 
the measurement result $z\in\{0,1\}^k$
for the first $k$ qubits
is
\begin{eqnarray}
p_z=\frac{1}{2^{N-1}}\sum_{x\in\{0,1\}^{N-1}}f(x),
\label{aho2}
\end{eqnarray}
where
\begin{eqnarray*}
f(x)&\equiv&\frac{1}{2^k}\sum_{y\in\{0,1\}^{N-1}}
(-1)^{z\cdot C_{1,...,k}(x0)+z\cdot C_{1,...,k}(y0)}
\delta_{C_{k+1,...,N}(x0),C_{k+1,...,N}(y0)}\\
&=&\frac{1}{2^k}\sum_{y\in S}
(-1)^{z\cdot C_{1,...,k}(x0)+z\cdot C_{1,...,k}(y0)},
\end{eqnarray*}
$C_{1,...,k}(x0)$ is the first $k$ bits of $C(x0)$,
$C_{k+1,...,N}(x0)$ is the last $N-k$ bits of $C(x0)$,
and $S\subseteq\{0,1\}^{N-1}$ is defined by
\begin{eqnarray*}
S&\equiv&\{y\in\{0,1\}^{N-1}~|~
C_{k+1,...,N}(x0)=C_{k+1,...,N}(y0)
\}.
\end{eqnarray*}
The subset $S$ can be obtained in polynomial time in the following
way:
\begin{itemize}
\item[1.]
Set $S=\{\}$.
\item[2.]
Repeat the following for all $\alpha\in\{0,1\}^k$.
\begin{itemize}
\item[2-1.]
Calculate $C^{-1}(\alpha C_{k+1,...,N}(x0))$.
\item[2-2.]
If $C^{-1}(\alpha C_{k+1,...,N}(x0))=y0$
for certain $y\in\{0,1\}^{N-1}$,
add $y$ to $S$.
\end{itemize}
\item[3.]
End.
\end{itemize}
From the construction,
$|S|\le2^k=poly(N)$.
Therefore, the value of $f(x)$ is exactly computable
in polynomial time for each $x$.
Furthermore, $|f(x)|\le 1$ because
\begin{eqnarray*}
|f(x)|\le\frac{1}{2^k}\sum_{y\in S}
|(-1)^{z\cdot C_{1,...,k}(x0)+z\cdot C_{1,...,k}(y0)}|
\le1.
\end{eqnarray*}
Let us generate random bit strings $x_1,...,x_T\in\{0,1\}^{N-1}$.
We then calculate 
\begin{eqnarray*}
\tilde{p}_z\equiv\frac{1}{T}\sum_{i=1}^Tf(x_i).
\end{eqnarray*}
Note that if we take $X_i=f(x_i)$, then 
$E(X_i)=p_z$, and therefore from
the Chernoff-Hoeffding bound, 
\begin{eqnarray*}
{\rm Pr}\Big[|\tilde{p}_z-p_z|\le\epsilon\Big]\ge
1-2e^{-\frac{T\epsilon^2}{2}}.
\end{eqnarray*}

For any polynomial $r$, let us take $\epsilon=\frac{1}{5\times 2^{2k}r}$.
Given $\{\tilde{p}_z\}_z$, define the probability
distribution $\{q_z\}_z$ by
\begin{eqnarray*}
q_z\equiv\frac{|\tilde{p}_z|}
{\sum_{z\in\{0,1\}^k}|\tilde{p}_z|}.
\end{eqnarray*}
Note that it is well defined, because
\begin{eqnarray*}
\sum_{z\in\{0,1\}^k}|\tilde{p}_z|>0,
\end{eqnarray*}
which is shown as follows:
\begin{eqnarray*}
\sum_{z\in\{0,1\}^k}|\tilde{p}_z|\ge
\sum_{z\in\{0,1\}^k}\tilde{p}_z\ge
\sum_{z\in\{0,1\}^k}(p_z-\epsilon)=
1-2^k\epsilon=1-\frac{1}{5\times 2^kr}>0.
\end{eqnarray*}
Furthermore, $\{q_z\}_z$ is obtained in polynomial time.
The distance between $\{p_z\}_z$ and $\{q_z\}_z$ is
\begin{eqnarray*}
\sum_{z\in\{0,1\}^k}|p_z-q_z|\le 5\times 2^{2k}\epsilon=\frac{1}{r}.
\end{eqnarray*}
It is shown as follows. First,
\begin{eqnarray*}
q_z&=&
\frac{|\tilde{p}_z|}
{\sum_{z\in\{0,1\}^k}|\tilde{p}_z|}\\
&\le&
\frac{p_z+\epsilon}
{1-2^k\epsilon}\\
&=&
(p_z+\epsilon)(1+2^k\epsilon+o(2^k\epsilon))\\
&=&
p_z+p_z2^k\epsilon+p_zo(2^k\epsilon)+\epsilon+2^k\epsilon^2
+\epsilon o(2^k\epsilon)\\
&\le&p_z+5\times2^k\epsilon.
\end{eqnarray*}
Second,
\begin{eqnarray*}
q_z&=&
\frac{|\tilde{p}_z|}
{\sum_{z\in\{0,1\}^k}|\tilde{p}_z|}\\
&\ge&
\frac{p_z-\epsilon}
{1+2^k\epsilon}\\
&=&
(p_z-\epsilon)(1-2^k\epsilon+o(2^k\epsilon))\\
&=&
p_z-p_z2^k\epsilon+p_zo(2^k\epsilon)
-\epsilon+2^k\epsilon^2-\epsilon o(2^k\epsilon)\\
&\ge&
p_z-5\times 2^k\epsilon.
\end{eqnarray*}

In this way, we have shown that we can calculate
in classical polynomial time the probability distribution $\{q_z\}_z$ that is close to $\{p_z\}_z$.
Our final task is to
show
that $\{q_z\}_z$ can be sampled in classical
polynomial time.
For simplicity,
let us assume that each $q_z$ is represented in the
$m$-bit binary:
\begin{eqnarray*}
q_z=\sum_{j=1}^m2^{-j}a_{z,j},
\end{eqnarray*}
where $a_{z,j}\in\{0,1\}$ for $j=1,2,...,m$.
(Otherwise, by polynomially 
increasing the size of $m$, we obtain exponentially
good approximations.)
The following algorithm samples
the probability distribution $\{q_z\}_z$. 
\begin{itemize}
\item[1.]
Randomly generate an $m$-bit string 
$(w_1,...,w_m)\in\{0,1\}^m$.
\item[2.]
Output $z\in\{0,1\}^k$ such that
\begin{eqnarray*}
\sum_{y<z}q_y\le\sum_{j=1}^m2^{-j}w_j<\sum_{y\le z}q_y,
\end{eqnarray*}
where $y<z$ and $y\le z$ mean the standard dictionary order.
(For example, for three bits,
$000<001<010<011<100<101<110<111$.)
\end{itemize}
In summary, we have shown that
$\{p_z\}_z$ can be sampled in polynomial time within a
$1/poly(N)$ L1-norm error.

\section{Proof of Theorem~\ref{BQPcomplete}}
\label{Sec:BQPcomp}
In this section, we show Theorem~\ref{BQPcomplete}.
Our proof consists of two parts. In the first
subsection, we show that
PDD-Max is BQP-hard. In the second subsection,
we show that PDD-Max is in BQP.

\subsection{BQP-hardness}
\begin{figure}[htbp]
\begin{center}
\includegraphics[width=0.35\textwidth]{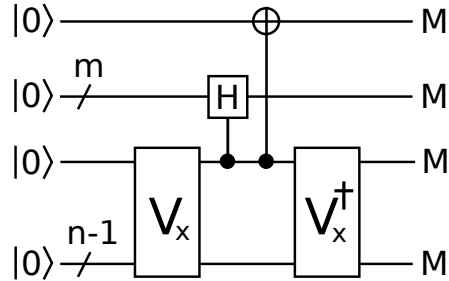}
\end{center}
\caption{
The circuit $U_1$ used to show the BQP-hardness of PDD-Max.
} 
\label{complete}
\end{figure}

%We first show that PDD-Max is BQP-hard.
Let $A=(A_{yes},A_{no})$ be a promise problem in BQP. 
It means
that there exists a uniform family $\{V_x\}_x$ of polynomial-size
quantum circuits such that if $x\in A_{yes}$ then
$V_x$ accepts with probability at least $1-2^{-r}$,
and if $x\in A_{no}$ then $V_x$ accepts with probability
at most $2^{-r}$,
where $r$ is any polynomial.
More precisely,
let $V_x$ be the 
polynomial-size quantum circuit, which acts on $n=poly(|x|)$ qubits,
corresponding to the instance
$x$. 
If we write $V_x|0^n\rangle$ as
\begin{eqnarray*}
V_x|0^n\rangle=\sqrt{\alpha}|0\rangle\otimes|\phi_0\rangle
+\sqrt{1-\alpha}|1\rangle\otimes|\phi_1\rangle
\end{eqnarray*}
with certain $(n-1)$-qubit states $|\phi_0\rangle$ and
$|\phi_1\rangle$, we have
$\alpha\ge1-2^{-r}$ when $x\in A_{yes}$,
and $\alpha\le 2^{-r}$ when $x\in A_{no}$.
Let us consider the circuit of Fig.~\ref{complete}.
We call it $U_1$. 
We also define
\begin{eqnarray*}
U_2\equiv H^{\otimes n+m+1}.
\end{eqnarray*}
We now show that deciding $x\in A_{yes}$ or $x\in A_{no}$ 
can be reduced to
a PDD-Max problem with $U_1$ and $U_2$, 
which means that PDD-Max is BQP-hard.
In fact, note that
\begin{eqnarray*}
U_1|0^{n+m+1}\rangle=
\sqrt{\alpha}|0\rangle\otimes|0^m\rangle\otimes
V_x^\dagger(|0\rangle\otimes|\phi_0\rangle)
+\sqrt{1-\alpha}|1\rangle\otimes|+\rangle^{\otimes m}\otimes
V_x^\dagger(|1\rangle\otimes|\phi_1\rangle).
\end{eqnarray*}
Let 
$p_z\equiv|\langle z|U_1|0^{n+m+1}\rangle|^2$ 
be the probability that we obtain $z\in\{0,1\}^{n+m+1}$
when we measure all qubits of $U_1|0^{n+m+1}\rangle$ 
in the computational basis.
When $x\in A_{yes}$, 
\begin{eqnarray*}
p_{0^{n+m+1}}=|\langle0^{n+m+1}|U_1|0^{n+m+1}\rangle|^2=
\alpha|\langle 0^n|V_x^\dagger(|0\rangle\otimes|\phi_0\rangle)|^2
=\alpha^2\ge(1-2^{-r})^2.
\end{eqnarray*}
When $x\in A_{no}$, 
\begin{eqnarray*}
p_{0y}&=&|\langle0y|U_1|0^{n+m+1}\rangle|^2=
\alpha\Big|\langle y|\Big[|0^m\rangle\otimes
V_x^\dagger(|0\rangle\otimes|\phi_0\rangle)\Big]
\Big|^2
\le \alpha\le 2^{-r}\\
p_{1y}&=&
|\langle1y|U_1|0^{n+m+1}\rangle|^2=
2^{-m}
(1-\alpha)|\langle y_{m+1},...,y_{n+m}|
V_x^\dagger(|1\rangle\otimes|\phi_1\rangle)|^2
\le 2^{-m}
\end{eqnarray*}
for any $y\in\{0,1\}^{n+m}$.
Let $q_z\equiv|\langle z|U_2|0^{n+m+1}\rangle|^2$,
where $z\in\{0,1\}^{n+m+1}$.
Since $U_2=H^{\otimes n+m+1}$, it is obvious that
$q_z=\frac{1}{2^{n+m+1}}$ for any $z$.
Therefore, when $x\in A_{yes}$,
\begin{eqnarray*}
|p_{0^{n+m+1}}-q_{0^{n+m+1}}|=
\Big|p_{0^{n+m+1}}-\frac{1}{2^{n+m+1}}\Big|\ge (1-2^{-r})^2-2^{-(n+m+1)},
\end{eqnarray*}
and when $x\in A_{no}$,
\begin{eqnarray*}
|p_z-q_z|&=&
\Big|p_z-\frac{1}{2^{n+m+1}}\Big|\\
&\le& p_z+\frac{1}{2^{n+m+1}}\\
&\le& \max(2^{-m},2^{-r})+2^{-(n+m+1)}
\end{eqnarray*}
for any $z\in\{0,1\}^{n+m+1}$.
In this way, we have shown that PDD-Max is BQP-hard.

\subsection{In BQP}
\label{Sec:inBQP}
We next show that PDD-Max is in BQP.
We show that the following BQP algorithm solves PDD-Max
with a $1/poly(N)$ completeness-soundness gap
(i.e., the gap of acceptance probabilities between the yes-instances
and the no-instances is lowerbounded by $1/poly(N)$):
\begin{itemize}
\item[1.]
Flip a fair coin $s\in\{0,1\}$. Generate $U_{s+1}|0^N\rangle$, 
and measure each qubit in the computational basis.
Let $z\in\{0,1\}^N$ be the measurement result.
\item[2.]
Repeat the following for $i=1,2,...,T$,
where $T$ is a polynomial of $N$ specified later.
\begin{itemize}
\item[2-a.]
Generate $U_1|0^N\rangle$, and measure all qubits
in the computational basis. If the result is $z$, then
set $X_i=1$. Otherwise, set $X_i=0$.
\end{itemize}
\item[3.]
Calculate 
\begin{eqnarray*}
\tilde{p}_z\equiv\frac{1}{T}\sum_{i=1}^TX_i.
\end{eqnarray*}
Note that $E(X_i)=1\times p_z+0\times (1-p_z)=p_z$,
where $p_z\equiv|\langle z|U_1|0^N\rangle|^2$.
Therefore, from the Chernoff-Hoeffding bound,
$|\tilde{p}_z-p_z|\le \epsilon$ with probability larger
than $1-2e^{-\frac{T\epsilon^2}{2}}$.
In a similar way, calculate the estimator $\tilde{q}_z$ 
of $q_z\equiv|\langle z|U_2|0^N\rangle|^2$.
From the Chernoff-Hoeffding bound,
$|\tilde{q}_z-q_z|\le \epsilon$ with probability larger
than $1-2e^{-\frac{T\epsilon^2}{2}}$.
\item[4.]
Calculate $|\tilde{p}_z-\tilde{q}_z|$.
If $|\tilde{p}_z-\tilde{q}_z|\ge a-\frac{a-b}{4}$, accept.
Otherwise, reject.
\end{itemize}

\if0
\begin{figure}[htbp]
\begin{center}
\includegraphics[width=0.4\textwidth]{swap.eps}
\end{center}
\caption{
The SWAP test between $|\psi\rangle$
and $|\phi\rangle$. ``SWAP" in the circuit means the
SWAP gate that exchanges two quantum states as
$|\psi\rangle\otimes|\phi\rangle\to|\phi\rangle\otimes|\psi\rangle$.
The first qubit is measured in the computational basis. 
Let $m\in\{0,1\}$ be the measurement result.
The expectation value of $(-1)^m$ 
is $|\langle\psi|\phi\rangle|^2$.
}
\label{swap}
\end{figure}
\fi

The intuitive idea of this algorithm is as follows.
By the definition of the PDD-Max,
if the answer of the PDD-Max is yes, there exists $z$
such that $|p_z-q_z|$ is large. In this case, the probability
of obtaining such $z$ in step 1 is large.
Therefore let us assume that we obtain such $z$ in step 1.
The step 1 is, in other words, the process to find a
candidate of the solution.
In steps 2 and 3, probabilities $p_z$ and $q_z$ are estimated
by using the Chernoff-Hoeffding bound.
Finally, in step 4, we check whether $|p_z-q_z|$ is indeed large,
and accept with high probability since it is actually large.
If the answer of the PDD-Max is no, on the other hand,
there is no
$z$ such that $|p_z-q_z|$ is large, and therefore in step 4 we do not
conclude that $|p_z-q_z|$ is large except for some failure probability. 
(Note that for general
$U_1$ and $U_2$, estimating $p_z$ and $q_z$ seems to require BQP power,
since it seems to be necessary to generate
$U_1|0^N\rangle$ and $U_2|0^N\rangle$.
We will see in the next section
that if $U_1$ and $U_2$ are restricted in FH$_2$, the estimation
of $p_z$ and $q_z$ can be done in classical polynomial time,
and therefore PDD-Max has a Merlin-Arthur system
with quantum
polynomial-time Merlin.)

Now let us give a more precise proof that
PDD-Max is in BQP.
First, let us consider the case when
the answer to PDD-Max is YES.
If $z$ obtained in step 1 satisfies $|p_z-q_z|\ge a$, and
if $\tilde{p}_z$ and $\tilde{q}_z$ 
calculated in steps 2 and 3 satisfy
$|\tilde{p}_z-p_z|\le\frac{a-b}{8}$ and
$|\tilde{q}_z-q_z|\le\frac{a-b}{8}$,
we definitely accept in step 4, because
\begin{eqnarray*}
a&\le& |p_z-q_z|\\
&\le&|p_z-\tilde{p}_z|+|\tilde{p}_z-q_z|\\
&\le&|p_z-\tilde{p}_z|+|\tilde{p}_z-\tilde{q}_z|+|\tilde{q}_z-q_z|\\
&\le&|\tilde{p}_z-\tilde{q}_z|+\frac{a-b}{4},
\end{eqnarray*}
and therefore
\begin{eqnarray*}
|\tilde{p}_z-\tilde{q}_z|\ge a-\frac{a-b}{4}.
\end{eqnarray*}

The probability $\eta$ of occurring such an event is
calculated to be
\begin{eqnarray*}
\eta
&\equiv&
{\rm Pr}[\mbox{$z$ obtained in step 1 satisfies $|p_z-q_z|\ge a$}]\\
&&\times
{\rm Pr}\Big[|\tilde{p}_z-p_z|\le\frac{a-b}{8}\Big]
\times{\rm Pr}\Big[|\tilde{q}_z-q_z|\le\frac{a-b}{8}\Big]\\
&=&
\Big(\sum_{z:|p_z-q_z|\ge a}\frac{p_z+q_z}{2}\Big)\times
{\rm Pr}\Big[|\tilde{p}_z-p_z|\le\frac{a-b}{8}\Big]
\times{\rm Pr}\Big[|\tilde{q}_z-q_z|\le\frac{a-b}{8}\Big]\\
&\ge& \frac{a}{2}
\Big(1-2e^{-\frac{T(a-b)^2}{128}}\Big)^2\\
&\ge& \frac{a}{2}(1-2e^{-k})^2,
\end{eqnarray*}
where we have taken $T\ge\frac{128k}{(a-b)^2}$ and $k$ is any polynomial
of $N$.
Therefore,
the acceptance probability $p_{acc}$ of our protocol is
lowerbounded as
\begin{eqnarray*}
p_{acc}\ge \eta\ge\frac{a}{2}(1-2e^{-k})^2\equiv\alpha.
\end{eqnarray*}

Next, let us consider the case when
the answer to PDD-Max is NO.
If $|\tilde{p}_z-p_z|\le\frac{a-b}{8}$
and $|\tilde{q}_z-q_z|\le\frac{a-b}{8}$,
which occurs with probability $\ge (1-2e^{-k})^2$,
we definitely reject,
because
\begin{eqnarray*}
|\tilde{p}_z-\tilde{q}_z|
&\le&|\tilde{p}_z-p_z|+|p_z-\tilde{q}_z|\\
&\le&|\tilde{p}_z-p_z|+|p_z-q_z|+|q_z-\tilde{q}_z|\\
&\le&\frac{a-b}{4}+b\\
&<&a-\frac{a-b}{4}.
\end{eqnarray*}
The rejection probability $p_{rej}$ is therefore lowerbounded
as
\begin{eqnarray*}
p_{rej}\ge (1-2e^{-k})^2.
\end{eqnarray*}
Therefore,
the acceptance probability $p_{acc}$ is
upperbounded as
\begin{eqnarray*}
p_{acc}= 1-p_{rej}\le
1-(1-2e^{-k})^2
=4e^{-k}-4e^{-2k}\equiv\beta.
\end{eqnarray*}
The completeness-soundness gap is
therefore 
\begin{eqnarray*}
\alpha-\beta=\frac{a}{2}(1-2e^{-k})^2-(4e^{-k}-4e^{-2k})\ge\frac{1}{poly(N)}
\end{eqnarray*}
for sufficiently large $k$,
which shows that PDD-Max is in BQP.

\section{Proof of Theorem~\ref{restriction_HC1Q}}
\label{Sec:FH_2}
In this section, we show Theorem~\ref{restriction_HC1Q}.
Before giving the proof, let us explain the intuitive idea.
In Sec.~\ref{Sec:inBQP}, we have seen that
PDD-Max is in BQP for general $U_1$ and $U_2$. 
In step 1 of the algorithm,
a candidate of the solution, i.e., $z$ such that
$|p_z-q_z|\ge a$, is obtained by doing quantum computing.
In step 2, another quantum computing is necessary to
estimate $p_z$ and $q_z$.
In the following
we will see that if $U_1$ and $U_2$ are the HC1Q model, 
estimations of $p_z$ and $q_z$ 
can be done in classical polynomial-time. 
Therefore, we can construct the following
Merlin-Arthur protocol with quantum polynomial-time Merlin:
\begin{itemize}
\item[1.]
Merlin does the step 1 of the protocol in Sec.~\ref{Sec:inBQP},
i.e., Merlin generates $z$ and sends it to Arthur.
\item[2.]
Arthur verifies $|p_z-q_z|$ is large.
\end{itemize}

Now let us show Theorem~\ref{restriction_HC1Q}.
%To understand our idea, we first 
%restrict $U_1$ and $U_2$ to be circuits in the form of 
%Fig.~\ref{simple2}(a). Similar results are obtained
%for other circuits in FH$_2$,
%such as IQP circuits and the Simon type ones.
%These generalizations are discussed in the next subsection.
Our Merlin-Arthur 
protocol with quantum polynomial-time Merlin runs as follows.
\begin{itemize}
\item[1.]
If Merlin is honest, he flips a fair coin $s\in\{0,1\}$. 
He next generates $U_{s+1}|0^N\rangle$,
and measures each qubit in the computational basis
to obtain the result $z\in\{0,1\}^N$.
He sends $z$ to Arthur. 
If Merlin is malicious, his computational ability is unbounded,
and what he sends to Arthur can be any $N$ bit string.
\item[2.]
Let $T$ be a polynomial of $N$ specified later.
Arthur generates $\{x^i\}_{i=1}^T$ and
$\{y^i\}_{i=1}^T$, 
where each of $x^i\in\{0,1\}^{N-1}$ and $y^i\in\{0,1\}^{N-1}$ 
is a uniformly and independently chosen random $N-1$ bit string.
\item[3.]
Arthur calculates 
\begin{eqnarray*}
\tilde{p}_z\equiv\frac{1}{T}\sum_{i=1}^T
f(x^i,y^i),
\end{eqnarray*}
where
\begin{eqnarray*}
f(x,y)\equiv
(-1)^{\sum_{j=1}^{N-1}[C_j(x0)\cdot z_j+C_j(y0)\cdot z_j]}
\delta_{C_N(x0),z_N}
\delta_{C_N(y0),z_N}.
\end{eqnarray*}
Here, 
\begin{eqnarray*}
C:\{0,1\}^N\ni w\mapsto C(w)\in\{0,1\}^N
\end{eqnarray*}
is the classical circuit in $U_1$,
\begin{eqnarray*}
x0\equiv(x_1,...,x_{N-1},0),
\end{eqnarray*}
$C_j(x0)$ is the $j$th bit of $C(x0)\in\{0,1\}^N$,
and $z_j$ is the $j$th bit of $z\in\{0,1\}^N$.
We call $\tilde{p}_z$ the estimator of $p_z$.
Note that $\tilde{p}_z$ can be calculated in classical $poly(N)$ time.
\item[4.]
In a similar way, Arthur calculates the estimator $\tilde{q}_z$
of $q_z$.
If 
\begin{eqnarray*}
|\tilde{p}_z-\tilde{q}_z|\ge a-\frac{a-b}{4}, 
\end{eqnarray*}
Arthur accepts.
Otherwise, he rejects.
\end{itemize}

Note that 
\begin{eqnarray}
p_z\equiv|\langle z|U_1|0^N\rangle|^2
=\frac{1}{2^{2(N-1)}}\sum_{x,y\in\{0,1\}^{N-1}}
f(x,y).
\label{aho}
\end{eqnarray}
Its derivation is given in Appendix~\ref{app1}.
If we set $X_i=f(x^i,y^i)$, we have
\begin{eqnarray*}
E(X_i)=\frac{1}{2^{2(N-1)}}\sum_{x,y\in\{0,1\}^{N-1}}f(x,y)=p_z.
\end{eqnarray*}
Therefore, from the Chernoff-Hoeffding bound, 
the estimator $\tilde{p}_z$ satisfies
$|\tilde{p}_z-p_z|\le\epsilon$ 
with probability $\ge1-2e^{-\frac{T\epsilon^2}{2}}$.
For $\tilde{q}_z$, a similar result holds.
In the above protocol,
the honest prover is enough to be quantum polynomial-time.

Now we show that the above protocol can verify PDD-Max.
The proof is similar to that for the BQP case given in 
Sec.~\ref{Sec:BQPcomp}.
First,
let us consider the case when the answer to 
PDD-Max is yes.
If the $z$ that Merlin sends to Arthur satisfies 
$|p_z-q_z|\ge a$,
and the estimators $\tilde{p}_z$ and $\tilde{q}_z$
that Arthur calculates satisfy 
$|\tilde{p}_z-p_z|\le\frac{a-b}{8}$
and $|\tilde{q}_z-q_z|\le\frac{a-b}{8}$,
Arthur definitely accepts,
because 
\begin{eqnarray*}
|\tilde{p}_z-\tilde{q}_z|&\ge&
|p_z-q_z|-|\tilde{p}_z-p_z|-|\tilde{q}_z-q_z|\\
&\ge&a-\frac{a-b}{4}.
\end{eqnarray*}
Taking the honest prover in step 1,
the probability $\eta$ of occurring such an event is
calculated to be
\begin{eqnarray*}
\eta
&\equiv&
{\rm Pr}[\mbox{Merlin obtains $z$ such that $|p_z-q_z|\ge a$}]\\
&&\times
{\rm Pr}\Big[|\tilde{p}_z-p_z|\le\frac{a-b}{8}\Big]
\times{\rm Pr}\Big[|\tilde{q}_z-q_z|\le\frac{a-b}{8}\Big]\\
&=&
\Big(\sum_{z:|p_z-q_z|\ge a}\frac{p_z+q_z}{2}\Big)\times
{\rm Pr}\Big[|\tilde{p}_z-p_z|\le\frac{a-b}{8}\Big]
\times{\rm Pr}\Big[|\tilde{q}_z-q_z|\le\frac{a-b}{8}\Big]\\
&\ge& \frac{a}{2}
\Big(1-2e^{-\frac{T(a-b)^2}{128}}\Big)^2
\ge \frac{a}{2}(1-2e^{-k})^2,
\end{eqnarray*}
where we have taken $T\ge\frac{128k}{(a-b)^2}$ and $k$ is any polynomial
of $N$.
Therefore,
the probability $p_{acc}$ that Arthur accepts in our protocol is
lowerbounded as
\begin{eqnarray*}
p_{acc}\ge \frac{a}{2}(1-2e^{-k})^2\equiv\alpha.
\end{eqnarray*}

Next, let us consider the case when
the answer to PDD-Max is no.
If $|\tilde{p}_z-p_z|\le\frac{a-b}{8}$
and $|\tilde{q}_z-q_z|\le\frac{a-b}{8}$,
which occurs with probability $\ge (1-2e^{-k})^2$,
Arthur definitely rejects,
because
\begin{eqnarray*}
|\tilde{p}_z-\tilde{q}_z|
&\le&|\tilde{p}_z-p_z|+|p_z-q_z|+|q_z-\tilde{q}_z|\\
&\le&b+\frac{a-b}{4}<a-\frac{a-b}{4}.
\end{eqnarray*}
Therefore,
the probability $p_{acc}$ that Arthur accepts in our protocol is
upperbounded as
\begin{eqnarray*}
p_{acc}\le 1-(1-2e^{-k})^2
=4e^{-k}-4e^{-2k}\equiv\beta.
\end{eqnarray*}
The completeness-soundness gap is
therefore 
\begin{eqnarray*}
\alpha-\beta=\frac{a}{2}(1-2e^{-k})^2-(4e^{-k}-4e^{-2k})\ge\frac{1}{poly(N)}
\end{eqnarray*}
for sufficiently large $k$,
which shows that PDD-Max has a Merlin-Arthur
system with quantum polynomial-time
Merlin.

{\bf Remarks}.
To conclude this section, we provide
some remarks. Note that the fact that $p_z$ and $q_z$ can be estimated in 
classical polynomial-time
seems to be a special property only for some circuits (such as those
in FH$_2$),
and we do not know how to do that for other general circuits.
(In particular, in Sec.~\ref{Sec:discussion2}, 
we will explain why the technique we use for FH$_2$ circuits
will not work
for other circuits, such as FH$_3$ circuits.)
Furthermore, note that what Arthur can do is to estimate 
$p_z$ within an additive error given $z$: he cannot sample $\{p_z\}_z$.
(If he can do it, he can solve factoring and Simon's problem,
for example.)

\section{Discussion}
\label{Sec:discussion}
\subsection{Generalizations to other circuits}
\label{Sec:discussion1}
In the proof of Theorem~\ref{restriction_HC1Q} (Sec.~\ref{Sec:FH_2}),
we have considered the HC1Q model.
Similar proofs hold for other circuits in FH$_2$.
For example, let us consider the IQP model.
For the IQP model, $f(x,y)$ is a complex-valued function,
but we can use the Chernoff-Hoeffding bound for complex
random variables introduced in Ref.~\cite{NestSchwarz}.

The other example is the HC$m$Q model.
We can show a similar result for it.
In fact, the probability $p_{s,t}$ of
obtaining the result $(s,t)\in\{0,1\}^{N-m}\times\{0,1\}^m$
is
\begin{eqnarray*}
p_{s,t}=\frac{1}{2^{2(N-m)}}
\sum_{x,y\in\{0,1\}^{N-m}}
(-1)^{C_{1,...,N-m}(x0^m)\cdot s
+C_{1,...,N-m}(y0^m)\cdot s}
\delta_{t,C_{N-m+1,...,N}(x0^m)}
\delta_{t,C_{N-m+1,...,N}(y0^m)},
\end{eqnarray*}
and therefore the 
Chernoff-Hoeffding bound argument can be used.
Here, $C_{1,...,N-m}(x0^m)$ is the first $N-m$ bits of 
$C(x0^m)$, and
$C_{N-m+1,...,N}(x0^m)$ is the last $m$ bits of
$C(x0^m)$.

Furthermore, we would like to point out that PDD-Max
can has a Merlin-Arthur system
with quantum polynomial-time Merlin for other circuits
outside of FH$_2$. The essential point of our proof is that
the output probability distributions, $p_z$ and $q_z$,
of $U_1$ and $U_2$ can be classically efficiently estimated.
This property itself is not restricted to circuits
in FH$_2$, and therefore we believe that our results should hold
for many other circuits outside of FH$_2$.

\subsection{FH$_3$}
\label{Sec:discussion2}
FH$_2$ circuits have nice structures such that
$p_z$ and $q_z$ can be estimated in classical polynomial-time.
We do not know how to do the same thing for other circuits such
as those in FH$_3$. The reason is as follows.
From a similar calculation given in Appendix~\ref{app1}, we can
show that the probability $p_z\equiv|\langle z|U|0^N\rangle|^2$
for an FH$_3$ circuit $U$ satisfies
\begin{eqnarray*}
p_z=\frac{1}{2^M}\sum_{x\in\{0,1\}^M}f(x)
\end{eqnarray*}
for certain $M=poly(N)$ and a certain function $f$.
However, in this case, we no longer have $|f(x)|\le1$, 
but $f$ is exponentially increasing as a function of $N$:
$f(x)=2^{s(N)}g(x)$, where
$|g(x)|\le1$ and $s$ is a polynomial.
Then, by using the Chernoff-Hoeffding bound,
we can obtain an estimator $\xi$ that satisfies
\begin{eqnarray*}
\Big|\xi-\frac{1}{2^M}\sum_x g(x)\Big|\le\epsilon
\end{eqnarray*}
with probability larger than $1-2e^{-\frac{T\epsilon^2}{2}}$.
The above inequality means
\begin{eqnarray*}
\Big|\xi-\frac{p_z}{2^{s(N)}}\Big|\le\epsilon,
\end{eqnarray*}
which also means
\begin{eqnarray*}
|2^{s(N)}\xi-p_z|\le2^{s(N)}\epsilon.
\end{eqnarray*}
In order to make $2^{s(N)}\epsilon=O(1/poly(N))$,
$\epsilon$ must be exponentially small, which means
that $T$ must be exponentially large.
Therefore, we cannot obtain a $1/poly(N)$-precision estimator
of $p_z$ in classical polynomial time.
Note that it is known that as long as we use $f$ as a black box,
the Chernoff-Hoeffding type bounds are optimal 
(up to some factors)~\cite{Canetti}.
This is the reason why our previous proof does not work for
other circuits.
%We do not know whether we can obtain similar results for
%other circuits than FH$_2$.

\acknowledgements
We thank Keisuke Fujii, Seiichiro Tani,
and Francois Le Gall for discussion.
TM is supported by JST ACT-I No.JPMJPR16UP, JST PRESTO,
and the Grant-in-Aid for Young Scientists (B) No.JP17K12637 of JSPS. 
YT is supported by the Program for Leading Graduate
Schools: Interactive Materials Science Cadet Program
and JSPS Grant-in-Aid for JSPS Research Fellow
No.JP17J03503.
HN is supported by the Grant-in-Aid for Scientific Research 
(A) Nos.26247016, 16H01705 and (C) No.16K00015 of JSPS.

\appendix

\section{Example}
\label{app:example}
\begin{figure}[htbp]
\begin{center}
\includegraphics[width=0.6\textwidth]{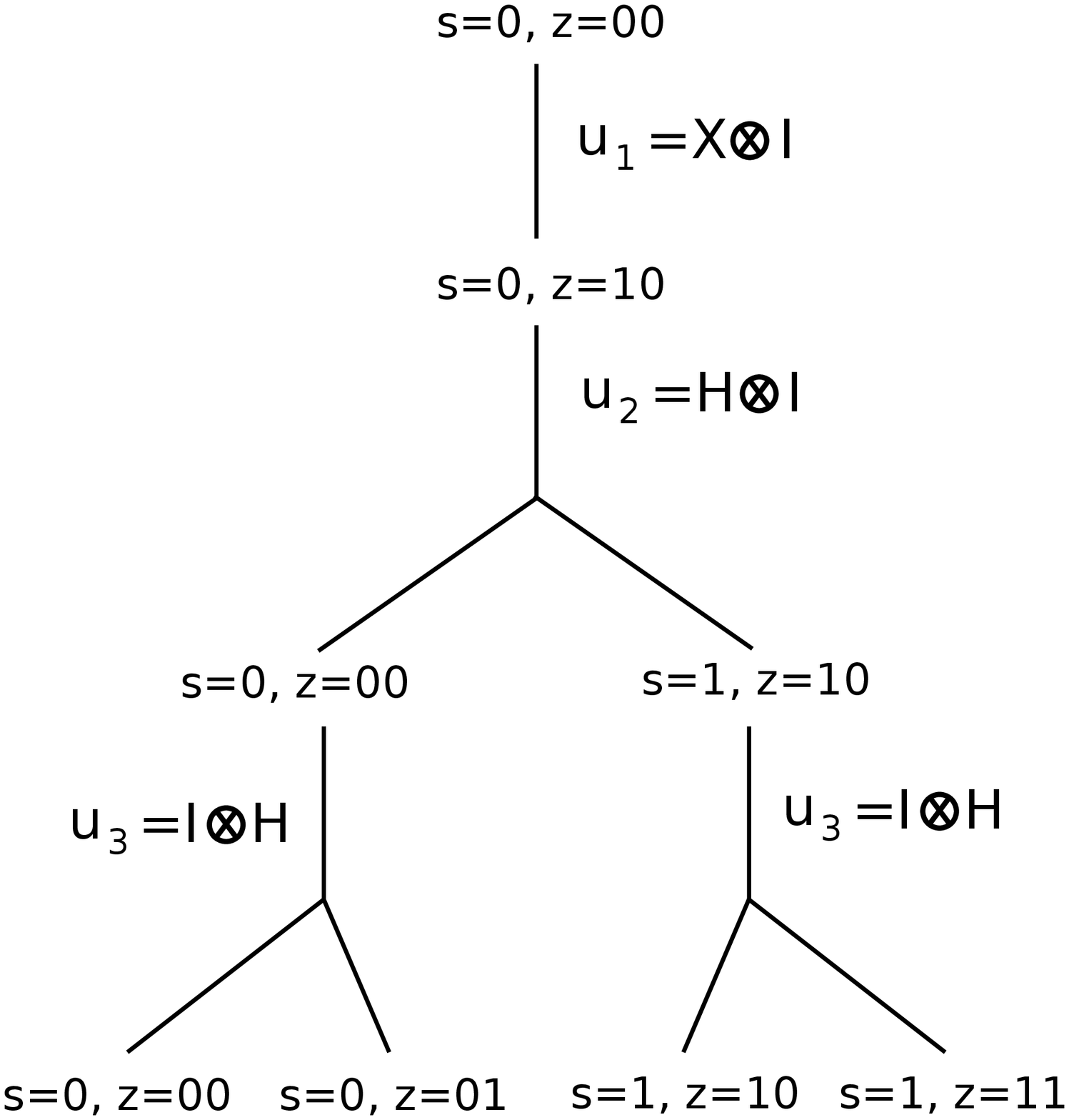}
\end{center}
\caption{
The computational tree of the non-deterministic algorithm.
} 
\label{tree}
\end{figure}

Let us consider the simple example where $n=2$,
$t=1$, and $U=u_1=X\otimes I$. 
In this case, 
$U'=u_3u_2u_1$, where
\begin{eqnarray*}
u_2&=&H\otimes I,\\
u_3&=&I\otimes H.
\end{eqnarray*}
The computational tree that represents the 
non-deterministic algorithm corresponding to $U'$ is given in Fig.~\ref{tree}.
Each Hadamard gate in $U'$ corresponds to a non-deterministic transition.
The bit string $z$ represents a basis state,
and the bit $s$ represents the sign of a basis state.
If we superpose $(-1)^{s(y)}|z(y)\rangle$ for all 
computational path $y\in\{0,1\}^2$,
we obtain
\begin{eqnarray*}
\sum_{y\in\{0,1\}^2}(-1)^{s(y)}|z(y)\rangle=
\Big[
(-1)^{0}|00\rangle
+(-1)^{0}|01\rangle
+(-1)^{1}|10\rangle
+(-1)^{1}|11\rangle
\Big]
=\sqrt{2^2}U'|0^2\rangle.
\end{eqnarray*}

\section{Derivation of Eq.~(\ref{aho2})}
\label{app2}
In this appendix, we show Eq.~(\ref{aho2}).
The initial state is $|0^N\rangle$.
We first apply $H^{\otimes (N-1)}\otimes I$.
Then the state becomes
\begin{eqnarray*}
\frac{1}{\sqrt{2^{N-1}}}\sum_{x\in\{0,1\}^{N-1}}|x\rangle|0\rangle.
\end{eqnarray*}
We apply the classical gate $C$ to obtain
\begin{eqnarray*}
\frac{1}{\sqrt{2^{N-1}}}\sum_{x\in\{0,1\}^{N-1}}|C(x0)\rangle.
\end{eqnarray*}
If we project this state onto 
$(H^{\otimes k}|z\rangle\langle z|H^{\otimes k})
\otimes I^{\otimes N-k}$, where
$z\in\{0,1\}^k$,
the state becomes
\begin{eqnarray*}
\frac{1}{\sqrt{2^{N-1}}}\frac{1}{\sqrt{2^k}}
\sum_{x\in\{0,1\}^{N-1}}
(-1)^{C_{1,...,k}(x0)\cdot z}
(H^{\otimes k}|z\rangle)\otimes|C_{k+1,...,N}(x0)\rangle.
\end{eqnarray*}
If we calculate its norm,
which is the probability $p_z$,
we obtain Eq.~(\ref{aho2}).

\section{Derivation of Eq.~(\ref{aho})}
\label{app1}
In this appendix, we show Eq.~(\ref{aho}).
The initial state is $|0^N\rangle$.
We first apply $H^{\otimes (N-1)}\otimes I$.
Then the state becomes
\begin{eqnarray*}
\frac{1}{\sqrt{2^{N-1}}}\sum_{x\in\{0,1\}^{N-1}}|x\rangle|0\rangle.
\end{eqnarray*}
We apply the classical gate $C$ to obtain
\begin{eqnarray*}
\frac{1}{\sqrt{2^{N-1}}}\sum_{x\in\{0,1\}^{N-1}}|C(x0)\rangle.
\end{eqnarray*}
We apply the final Hadamard $H^{\otimes (N-1)}\otimes I$ to obtain
\begin{eqnarray*}
\frac{1}{2^{N-1}}\sum_{x,y\in\{0,1\}^{N-1}}
(-1)^{\sum_{j=1}^{N-1}C_j(x0)\cdot y_j}|y\rangle|C_N(x0)\rangle
=U_1|0^N\rangle,
\end{eqnarray*}
where $C_j(x0)$ is the $j$th bit of $C(x0)$ and $y_j$ is
the $j$th bit of $y$.
Therefore
\begin{eqnarray*}
\langle z|U_1|0^N\rangle=
\frac{1}{2^{N-1}}\sum_{x\in\{0,1\}^{N-1}}
(-1)^{\sum_{j=1}^{N-1}C_j(x0)\cdot z_j}\delta_{C_N(x0),z_N},
\end{eqnarray*}
which finally gives Eq.~(\ref{aho})
\begin{eqnarray*}
p_z=|\langle z|U_1|0^N\rangle|^2=\frac{1}{2^{2(N-1)}}
\sum_{x,y\in\{0,1\}^{N-1}}f(x,y).
\end{eqnarray*}


\begin{thebibliography}{99}

\bibitem{FH}
Y. Shi,
Quantum and classical tradeoffs.
Theoretical Computer Science {\bf344}, 335 (2005).
\href
{https://doi.org/10.1016/j.tcs.2005.03.053}
{DOI:10.1016/j.tcs.2005.03.053}

\bibitem{Simon}
D. R. Simon,
On the power of quantum computation.
Proceedings of the
35th Annual Symposium on Foundations of Computer Science
(FOCS 1994), p.116 (1994).
\href{https://doi.org/10.1137/S0097539796298637}
{DOI:10.1137/S0097539796298637}

\bibitem{Shor}
P. Shor,
Algorithms for quantum computation: discrete logarithms and factoring.
Proceedings of the
35th Annual Symposium on Foundations of Computer Science 
(FOCS 1994), p.124 (1994).
\href{https://doi.org/10.1109/SFCS.1994.365700}
{DOI:10.1109/SFCS.1994.365700}

\bibitem{BJS}
M. J. Bremner, R. Jozsa, and D. J. Shepherd,
Classical simulation of commuting quantum computations implies
collapse of the polynomial hierarchy.
Proc. R. Soc. A {\bf467}, 459 (2011).
\href{https://doi.org/10.1098/rspa.2010.0301}
{DOI:10.1098/rspa.2010.0301}

\bibitem{BMS}
M. J. Bremner, A. Montanaro, and D. J. Shepherd,
Average-case complexity versus approximate simulation of commuting
quantum computations.
Phys. Rev. Lett. {\bf117}, 080501 (2016).
\href{https://doi.org/10.1103/PhysRevLett.117.080501}
{DOI:10.1103/PhysRevLett.117.080501}

\bibitem{TD}
B. M. Terhal and D. P. DiVincenzo,
Adaptive quantum computation,
constant depth quantum circuits and Arthur-Merlin
games.
Quant. Inf. Comput. {\bf4}, 134 (2004).
\href{https://doi.org/10.26421/QIC4.2}
{DOI:10.26421/QIC4.2}

\bibitem{AA}
S. Aaronson and A. Arkhipov,
The computational complexity of linear optics.
Theory of Computing {\bf9}, 143 (2013).
\href{http://doi.org/10.1145/1993636.1993682}
{DOI:10.1145/1993636.1993682}

\bibitem{KL}
E. Knill, and R. Laflamme, 
Power of one bit of quantum information.
Phys. Rev. Lett. {\bf 81}, 5672 (1998).
\href{https://doi.org/10.1103/PhysRevLett.81.5672}
{DOI:10.1103/PhysRevLett.81.5672}



\bibitem{MFF}
T. Morimae, K. Fujii, and J. F. Fitzsimons,
Hardness of classically simulating the one clean qubit model.
Phys. Rev. Lett. {\bf 112}, 130502 (2014).
\href{https://doi.org/10.1103/PhysRevLett.112.130502}
{DOI:10.1103/PhysRevLett.112.130502}


\bibitem{M}
T. Morimae,
Hardness of classically sampling one clean qubit model with
constant total variation distance error.
Phys. Rev. A {\bf96}, 040302(R) (2017).
\href{https://doi.org/10.1103/PhysRevA.96.040302}
{DOI:10.1103/PhysRevA.96.040302}



\bibitem{Kobayashi}
K. Fujii, H. Kobayashi, T. Morimae, H. Nishimura, S. Tamate,
and S. Tani,
Impossibility of classically simulating one-clean-qubit computation.
Phys. Rev. Lett. {\bf120}, 200502 (2018).
\href{https://doi.org/10.1103/PhysRevLett.120.200502}
{DOI:10.1103/PhysRevLett.120.200502}


\bibitem{KobayashiICALP}
K. Fujii, H. Kobayashi, T. Morimae, H. Nishimura, S. Tamate,
and S. Tani,
Power of quantum computation with few clean qubits.
Proceedings of 43rd International Colloquium on Automata,
Languages, and Programming (ICALP 2016), p.13:1.
\href{http://doi.org/10.4230/LIPIcs.ICALP.2016.13}
{DOI:10.4230/LIPIcs.ICALP.2016.13}

\bibitem{BU}
B. Fefferman and C. Umans,
The power of quantum Fourier sampling.
arXiv:1507.05592

\bibitem{AFK}
A. Bouland, J. F. Fitzsimons, and D. E. Koh,
Quantum advantage from conjugated Clifford circuits.
%arXiv:1709.01805
Proceedings of the 33rd Computational Complexity Conference (CCC2018).
\href{https://doi.org/10.4230/LIPIcs.CCC.2018.21}
{DOI:10.4230/LIPIcs.CCC.2018.21}

\bibitem{random}
A. Bouland, B. Fefferman, C. Nirkhe, and U. Vazirani,
On the complexity and verification of
quantum random circuit sampling.
%arXiv:1803.04402
Nat. Phys. 2018
\href{https://doi.org/10.1038/s41567-018-0318-2}
{DOI:10.1038/s41567-018-0318-2}

\bibitem{AharonovVazirani}
D. Aharonov and U. Vazirani,
Is quantum mechanics falsifiable? A computational perspective
on the foundations of quantum mechanics.
arXiv:1206.3686

\bibitem{MNS}
T. Morimae, D. Nagaj, and N. Schuch,
Quantum proofs can be verified using only single-qubit
measurements.
Phys. Rev. A {\bf93}, 022326 (2016).
\href{https://doi.org/10.1103/PhysRevA.93.022326}
{DOI:10.1103/PhysRevA.93.022326}

\bibitem{posthoc}
J. F. Fitzsimons, M. Hajdu{\v s}ek, and T. Morimae,
Post hoc verification of quantum computation.
Phys. Rev. Lett. {\bf120}, 040501 (2018).
\href{https://doi.org/10.1103/PhysRevLett.120.040501}
{DOI:10.1103/PhysRevLett.120.040501}


\bibitem{FK}
J. F. Fitzsimons and E. Kashefi,
Unconditionally verifiable blind computation.
Phys. Rev. A {\bf96}, 012303 (2017).
\href{https://doi.org/10.1103/PhysRevA.96.012303}
{DOI:10.1103/PhysRevA.96.012303}

\bibitem{Aharonov}
D. Aharonov, M. Ben-Or, E. Eban, and U. Mahadev,
Interactive proofs for quantum computations.
arXiv:1704.04487

\bibitem{HM}
M. Hayashi and T. Morimae,
Verifiable measurement-only blind quantum computing
with stabilizer testing.
Phys. Rev. Lett. {\bf115}, 220502 (2015).
\href{https://doi.org/10.1103/PhysRevLett.115.220502}
{DOI:10.1103/PhysRevLett.115.220502}


\bibitem{Broadbent}
A. Broadbent,
How to verify quantum computation.
Theory of Computing {\bf14}, 1 (2018).
\href{https://doi.org/10.4086/toc.2018.v014a011}
{DOI:10.4086/toc.2018.v014a011}

\bibitem{Andru_review}
A. Gheorghiu, T. Kapourniotis, and E. Kashefi,
Verification of quantum computation: an overview of existing
approaches.
arXiv:1709.06984

\bibitem{MattMBQC}
M. McKague,
Interactive proofs for BQP via self-tested graph states.
Theory of Computing {\bf12}, 1 (2016).
\href{https://doi.org/10.4086/toc.2016.v012a003}
{DOI:10.4086/toc.2016.v012a003}

\bibitem{Ji}
Z. Ji,
Classical verification of quantum proofs.
Proceedings of the 48th annual ACM symposium on Theory
of Computing (STOC 2016) p.885 (2016).
\href{https://doi.org/10.1145/2897518.2897634}
{DOI:10.1145/2897518.2897634}

\bibitem{RUV}
B. W. Reichardt, F. Unger, and U. Vazirani,
Classical command of quantum systems.
Nature {\bf496}, 456 (2013).
\href{https://doi.org/10.1038/nature12035}
{DOI:10.1038/nature12035}

\bibitem{defIP}
IP is the class of promise problems verified by an interactive proof system
with a BPP verifier and a prover with unlimited computational ability.

\bibitem{AharonovGreen}
D. Aharonov and A. Green,
A quantum inspired proof of 
${\rm P}^{\# {\rm P}}\subseteq {\rm IP}$.
arXiv:1710.09078

\bibitem{Mahadev}
U. Mahadev, Classical verification of quantum computations.
arXiv:1804.01082

\bibitem{BV}
E. Bernstein and U. Vazirani,
Quantum complexity theory. 
SIAM Journal on Computing {\bf26}, 1411 (1997).
\href{https://doi.org/10.1137/S0097539796300921}
{DOI:10.1137/S0097539796300921}

\bibitem{MattFourier}
M. McKague,
Interactive proofs with efficient quantum prover for
recursive Fourier sampling.
Chicago Journal of Theoretical Computer Science
{\bf6}, 1 (2012).
\href{https://doi.org/10.4086/cjtcs.0006}
{DOI:10.4086/cjtcs.0006}

\bibitem{Tommaso}
T. F. Demarie, Y. Ouyang, and J. F. Fitzsimons,
Classical verification of quantum circuits containing few basis
changes.
Phys. Rev. A {\bf97}, 042319 (2018).
\href{https://doi.org/10.1103/PhysRevA.97.042319}
{DOI:10.1103/PhysRevA.97.042319}

\bibitem{LeGall}
F. Le Gall, T. Morimae, H. Nishimura, and Y. Takeuchi,
Interactive proofs with polynomial-time quantum prover for
computing the order of solvable groups.
arXiv:1805.03385

\bibitem{defMA}
Merlin-Arthur (MA) is the class of promise problems
that can be verified by a BPP verifier and
a prover whose computational ability is unlimited.
In MA, the prover first sends a bit string to the verifier,
and the verifier then does the computing for the verification.


\bibitem{keep}
Note that the $(n+1)$-qubit Toffoli gate in Fig.~\ref{proof2}
can be implemented with a polynomially many Toffoli gates by
using ancilla qubits~\cite{Barenco}. It is known that
ancilla qubits do not need to be initialized~\cite{Barenco}.
Therefore, we can use $|+\rangle$ states
as ancilla qubits, which keeps the circuit of Fig.~\ref{proof2}
in the form of Fig.~\ref{simple2}(a).


\bibitem{Barenco}
A. Barenco, C. H. Bennett, R. Cleve, D. P. DiVincenzo, 
N. Margolus, P. Shor, T. Sleator, J. A. Smolin, H. Weinfurter, 
Elementary gates for quantum computation. 
Phys. Rev. A {\bf52} 3457 (1995).
\href{https://doi.org/10.1103/PhysRevA.52.3457}
{DOI:10.1103/PhysRevA.52.3457}

\bibitem{NQP}
L. M. Adleman, J. DeMarrais, and M. A. Huang, Quantum 
Computability. SIAM J. Comput. {\bf26} 1524 (1997).
\href{https://doi.org/10.1137/S0097539795293639}
{DOI:10.1137/S0097539795293639}

\bibitem{NestSchwarz}
M. Schwarz and M. Van den Nest,
Simulating quantum circuits with sparse output distributions.
arXiv:1310.6749

\bibitem{Canetti}
R. Canetti, G. Even, and O. Goldreich,
Lower bounds for sampling algorithms for estimating the average.
Information Processing Letters {\bf53}, 17 (1995).



\end{thebibliography}
\end{document}